\documentclass[12pt]{article}

\usepackage{epsfig}

\def\beq{\begin{equation}}
\def\eeq{\end{equation}}
\def\bea{\begin{eqnarray}}
\def\beaa{\begin{eqnarray*}}
\def\eea{\end{eqnarray}}
\def\eeaa{\end{eqnarray*}}
\def\bq{\begin{quote}}
\def\eq{\end{quote}}
\def\gappeq{\mathrel{\rlap {\raise.5ex\hbox{$>$}}
{\lower.5ex\hbox{$\sim$}}}}
\def\lappeq{\mathrel{\rlap{\raise.5ex\hbox{$<$}}
{\lower.5ex\hbox{$\sim$}}}}

{\end{list}}
\newcounter{enumct}

\newcommand{\captive}[1]{\rule{5mm}{0mm}%
\begin{minipage}{150mm}\caption[small]{#1}\end{minipage}}
 
\begin{document}
 
\sloppy

\pagestyle{empty}

\rightline{CERN-TH/99-350}
\rightline{hep-ph/9911440}
\begin{center}
{\LARGE\bf Possible Accelerators $@$ CERN Beyond the LHC }
\\[10mm]
{\Large John Ellis} \\[3mm]
{\it Theoretical Physics Division, CERN}\\[1mm]
{\it CH - 1211 Geneva 23}\\[1mm]
{\it E-mail: john.ellis@cern.ch}\\[20mm]

{\bf Abstract}\\[1mm]
\begin{minipage}[t]{140mm}
The physics and world-wide accelerator context for possible accelerator
projects at CERN after the LHC are reviewed, including the expectation
that an $e^+ e^-$ linear collider in the TeV energy range will be built
elsewhere. Emphasis is laid on the Higgs boson, supersymmetry and
neutrino oscillations as benchmarks for physics after the LHC. The default
option for CERN's next major project may be the CLIC multi-TeV $e^+ e^-$
collider project. Also interesting is the option of a three-step scenario
for muon storage rings, starting with a neutrino factory, continuing with
one or more Higgs factories, and culminating in a $\mu^+ \mu^-$ collider
at the high-energy frontier. \end{minipage}\\[5mm]

\rule{160mm}{0.4mm}
{\it To appear in the \\
Proceedings of the Workshop on the Development of Future Linear
Electron-Positron Colliders \\
for Particle Physics Studies and for Research using Free-Electron
Lasers \\
Lund, 23 - 26 September 1999}

\end{center}

\section{The Context}

By comparison with other high-energy physics laboratories, CERN is fortunate to
have an exciting physics programme beyond the year 2005 already approved
and under
construction, centred on the LHC. However, the time scales for the R\&D,
approval
and construction of major new accelerators are very long: the first LEP physics
study started in 1975~\cite{LEPYB}, 14 years before the first data, and
the first LHC
physics
study was in 1984~\cite{Lausanne}. Therefore, it is already time to be
thinking what CERN might
do for an encore after (say) ten years of physics with the LHC. Although
necessary, extrapolation to the likely physics agenda beyond 2015 is foolhardy,
since several major accelerators will be providing cutting-edge data during the
intervening period, and we do not know what they will find. (Otherwise, it
would
not be research, would it?) Nevertheless, we should try to set the apr\`es-LHC
era in context by surveying the ground that these intervening accelerators will
cover~\cite{EKR}, even if our crystal ball does not reveal what they will
find there.

LEP operation will terminate in 2000, after providing sensitivity to Higgs
masses
below about 110 GeV. The current lower limit from the  data of an individual
LEP experiment reaches about 106~GeV, as seen in Fig. 1~\cite{LEPC},
and a combined analysis of the full 1999 data might
increase the sensitivity to about 109 GeV. The most optimistic projection for
2000 that I have seen would extend this to about 113 GeV. Clearly, the
overall
picture changes if LEP discovers the Higgs boson. However, the precision
electroweak data and supersymmetric models independently suggest that $m_H
\lappeq$ 200 GeV, as seen in Fig. 2~\cite{LEPEWWG}, in which case the
programme of
exploring in
detail the properties of the Higgs boson is already well posed, just as the LEP
programme was outlined before the discovery of the $W^\pm$ and $Z^0$.

\begin{figure}
\begin{center}
\mbox{\epsfig{file=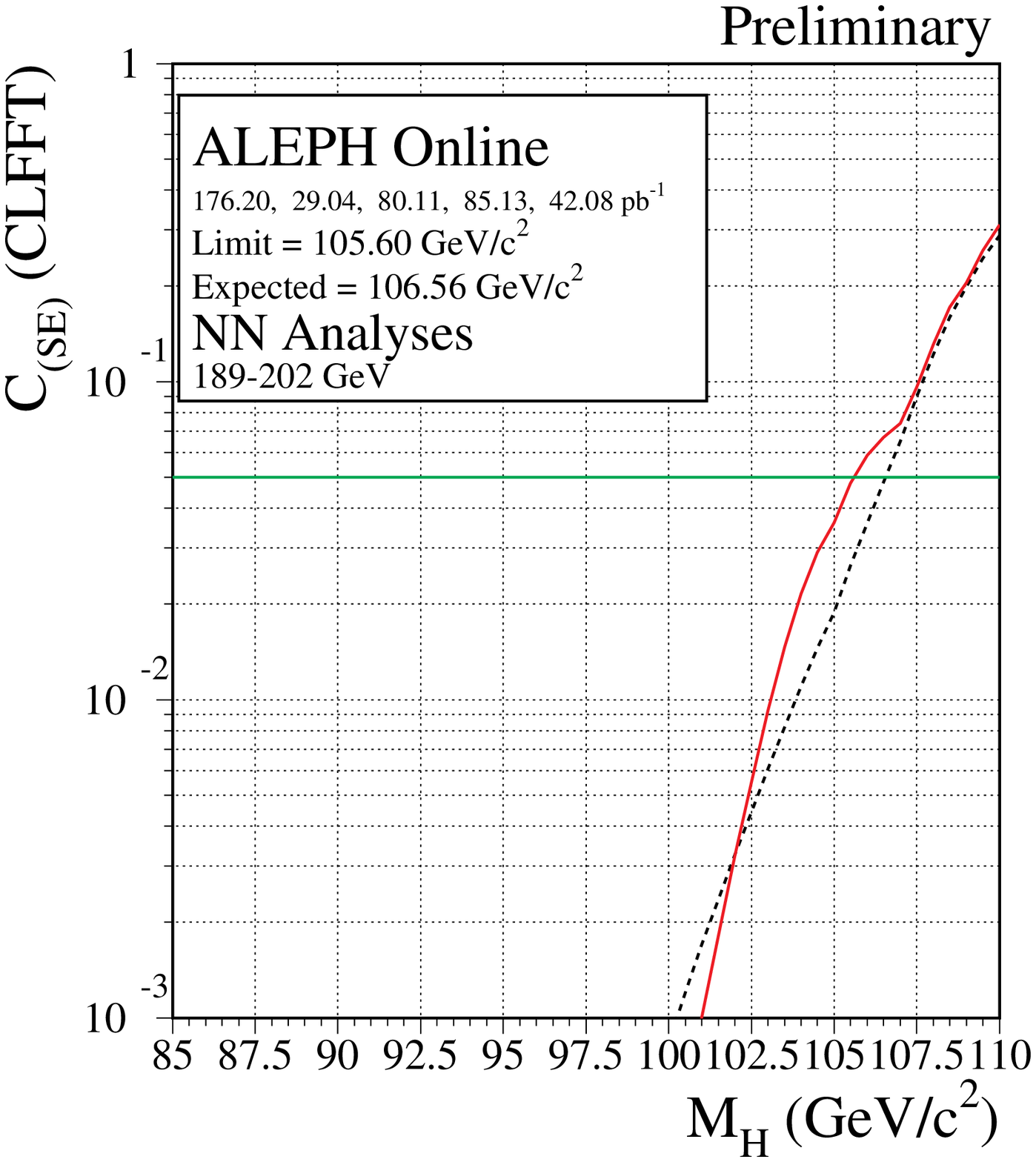,width=79mm}}
\end{center}
\captive{\it Preliminary lower limit on the Standard Model Higgs mass
obtained by the ALEPH collaboration~\cite{LEPC,ALEPH}.}
\label{figure}
\end{figure}

\begin{figure}
\begin{center}
\mbox{\epsfig{file=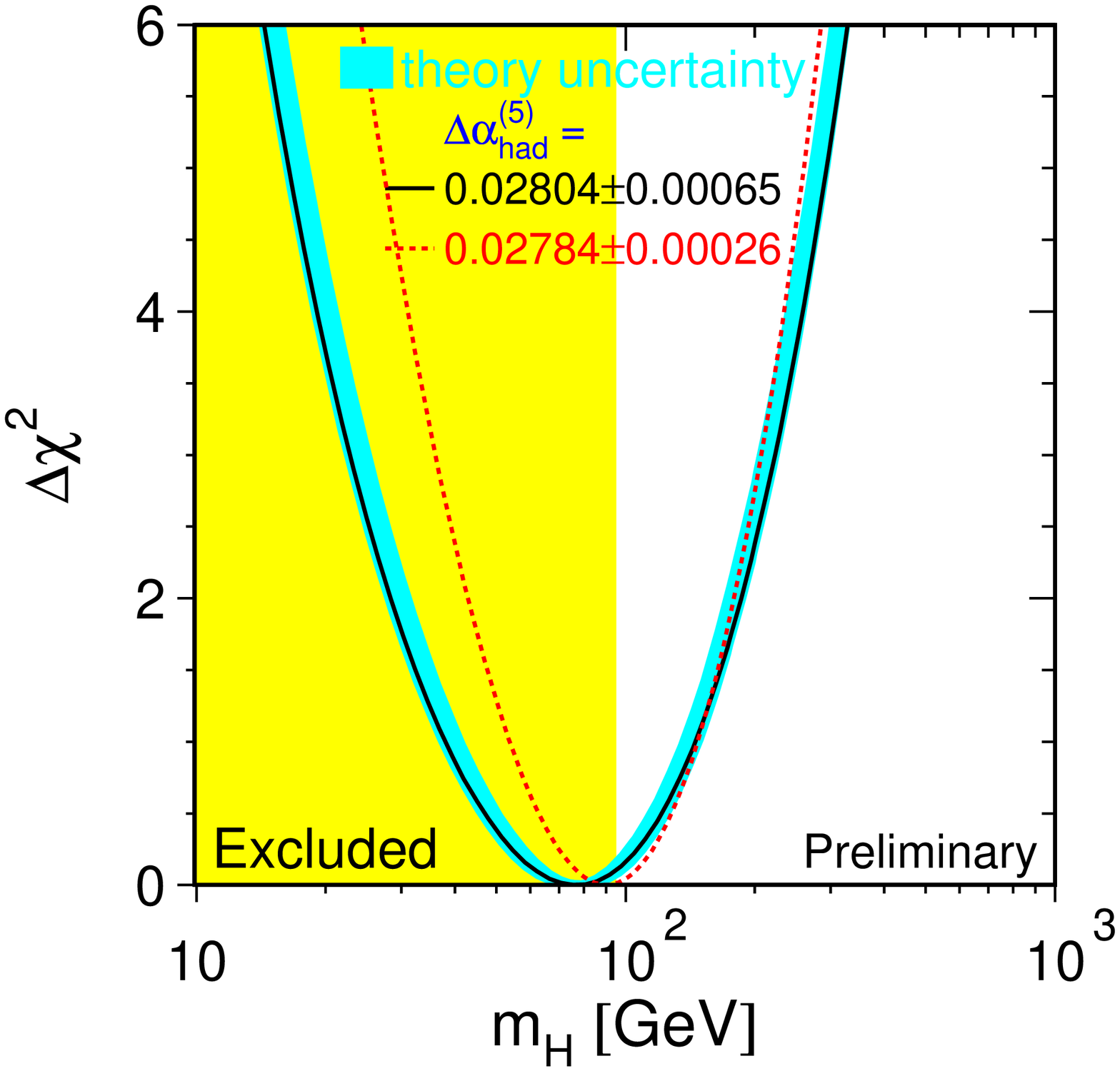,width=8cm}}
\end{center}
\captive{\it Estimate of the Higgs boson mass obtained from precision
electroweak data~\cite{LEPEWWG}.}
\end{figure}

CDF and
D$\phi$ at the FNAL Tevatron collider have a
chance to find the Higgs
boson before the LHC in its next run starting in 2001, as seen in Fig.
3~\cite{Carena}. This
figure is based on theoretical assessments of the capabilities of the Tevatron
detectors, and the experiments may fare better or worse. However, taken at face
value, it seems that the Tevatron detectors would need more than 5 or even 10
pb$^{-1}$ to explore masses beyond LEP's reach. Will these be available for the
LHC's scheduled start in 2005? FNAL's window of opportunity will extend
somewhat
beyond LHC start-up, since ATLAS and CMS  will take some time  to accumulate
the luminosity needed to explore the difficult region $M_H \lappeq$ 130
GeV~\cite{TDR}.

\begin{figure}
\begin{center}
\mbox{\epsfig{file=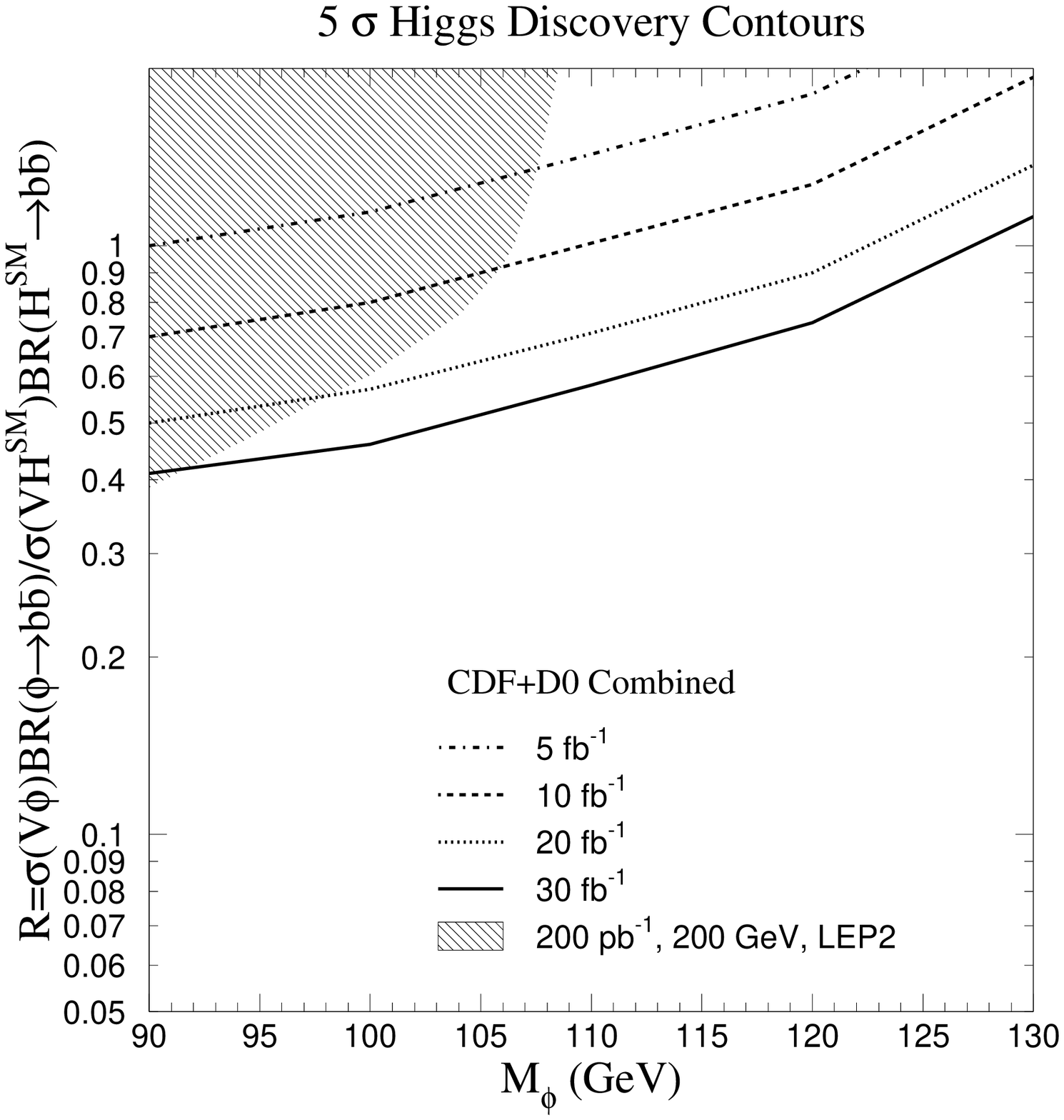,width=8cm}}
\end{center}
 \captive{\it Comparison of the estimated physics reaches for
Higgs searches at LEP~2 and the FNAL Tevatron collider~\cite{Carena},
as a function of Higgs mass and collider luminosity.}
\end{figure}

CDF~\cite{CDF} and the
OPAL~\cite{OPAL} and
ALEPH~\cite{ALEPH} experiments
at LEP together measure $\sin 2\beta =
0.91 \pm 0.35$ in $B\rightarrow J/\psi K_s$ decays.
The $B$ factories PEP-II (with BaBar) and KEK-B (with Belle) have already
started
operation, to be followed soon by HERA (with HERA-B) and CESR (with CLEO
3). All being well, the first
accurate measurements of CP violation in the $B$ sector should emerge in
2000, and
many more will follow. There are also ample $B$-physics opportunities for
CDF and
D$\phi$, which will be the first to explore CP violation in the $B_s$ system.
These will be followed by the LHC experiments, particularly LHCb, and
possibly by
BTeV. My crystal ball gets cloudy at this point: will any non-Standard-Model
physics reveal itself in $B$ decays? If so, there may be more heavy-flavour
physics to pursue~\cite{Nir}.

A promising new area of exploration has been opened by the strong
indications for
neutrino oscillations found by Super-Kamiokande~\cite{SK} et
al.~\cite{al}, and several
long-baseline
neutrino projects are underway. K2K has started taking data, and will be
able to
measure $\nu_\mu$ disappearance in much of the region of atmospheric-neutrino
parameter space favoured by Super-Kamiokande~\cite{K2K}. Starting in 2001,
KamLAND~\cite{KamLAND} will
explore the large-mixing-angle (LMA) MSW solution of the solar-neutrino
problem.
In 2003/2004, MINOS will start exploring $\nu_\mu$ disappearance, the NC/CC
ratio and other oscillation signatures in the FNAL NuMI beam~\cite{MINOS}.
The CERN-Gran Sasso
beam is planned to start providing opportunities in 2005 to look for
$\nu_\mu\rightarrow\nu_\tau$ oscillations via $\tau$
production~\cite{CNGS}.

What of the LHC? As seen in Fig. 4, it will discover the Standard-Model Higgs
boson (if this has not been done already), but this may take some
time~\cite{TDR}. It will
also be able to discover Higgs bosons in the minimal supersymmetric
extension of
the Standard Model (MSSM), though perhaps not all of them. It will also find
supersymmetry (if this has not been done already), establish much of the
sparticle
spectrum, as displayed in Table 1~\cite{LHCsusy}, and measure some
distinctive spectral
features, as seen in Fig. 5. To baseline the subsequent discussion, we surmise
that the LHC will not only discover the Higgs boson, but also measure its mass
with a precision between 0.1 \% and 1 \%~\cite{TDR}. However, it will only
be able to
observe a couple of Higgs decay modes. Within the context of the MSSM, the LHC
will have found many sparticles, but perhaps not the heavier Higgs bosons and
weakly-interacting sparticles such as sleptons and
charginos~\cite{LHCsusy}. The spectroscopic
measurements will not enable the underlying MSSM parameters to be strongly
over-constrained.

\begin{table}[h]
\caption{\it The LHC as `Bevatrino': Sparticles detectable~\cite{LHCsusy}
at
five selected points in supersymmetric parameter space are denoted by +}
\begin{center}
\begin{tabular}{|c|c|c|c|c|c|c|c|c|c|c|c|c|}   \hline
&&&&&&&&&&&& \\
 & $h$ & $H/A$ & $\chi^0_2$ & $\chi^0_3$ & $\chi^-_1$  &$\chi^\pm_1$ & $\chi^\pm_2$ &
$\tilde q$ & $\tilde b$ & $\tilde t$ & $\tilde g$ & $\tilde\ell$\\ \hline 
1 & + && + &&&&& + & + & + & + &\\  \hline
2 & + &&+ &&&&& + & + & + & + & \\  \hline
3 & + & + & + &&&+ && + & + && + &\\  \hline
4 & + && + & + & + & + & + & + &&& + & \\ \hline
5 & + && + &&&&& + & + & + & + & +  \\ \hline
\end{tabular}
\end{center}
\end{table}

\begin{figure}
\begin{center}
\mbox{\epsfig{file=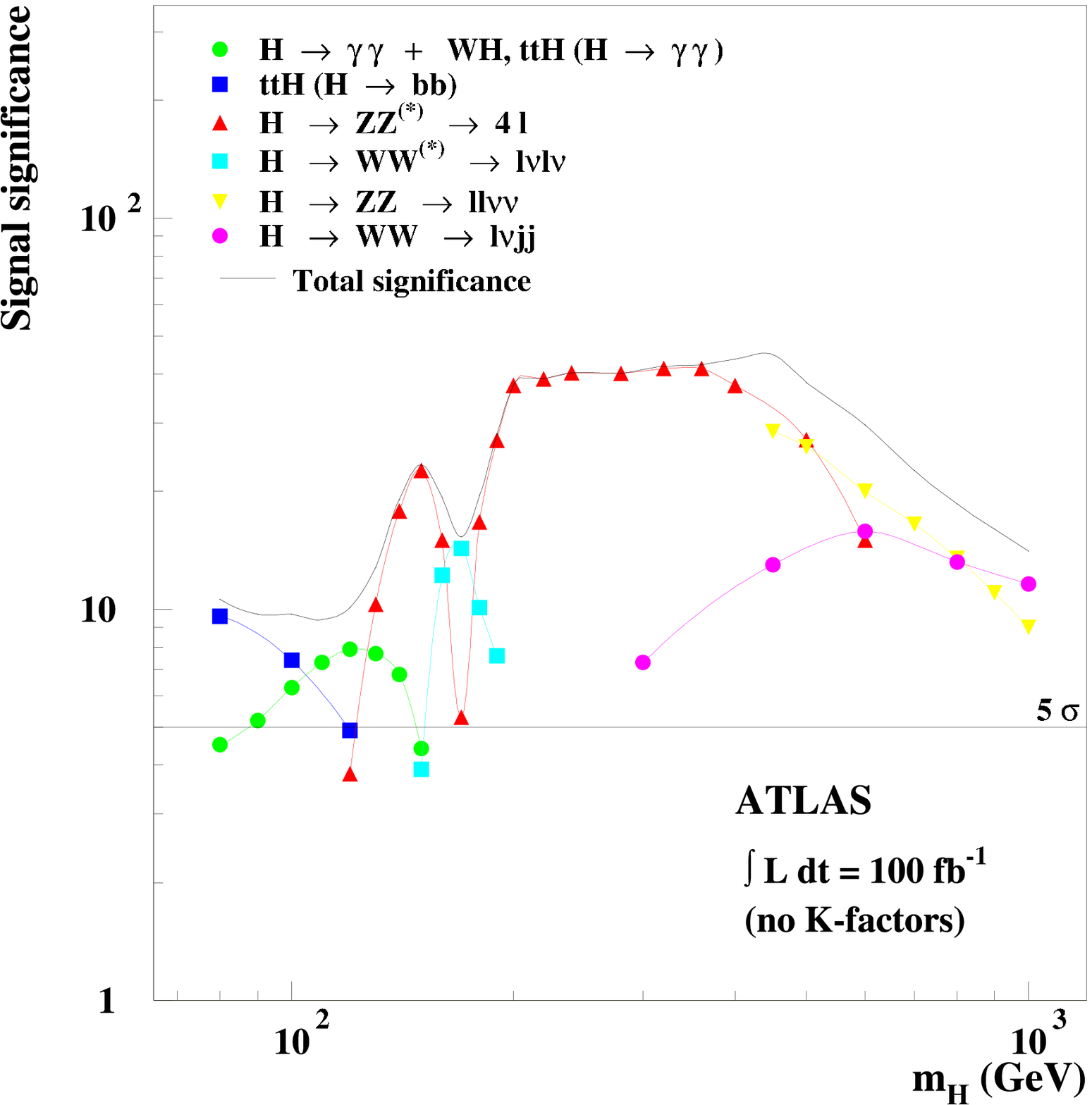,width=8cm}}
\end{center}
 \captive{\it Estimated significance of the possible
Higgs detection at the LHC in various channels, as
a function of the assumed value of the
Higgs mass~\cite{TDR}.}
\end{figure}

\begin{figure}
\begin{center}
\mbox{\epsfig{file=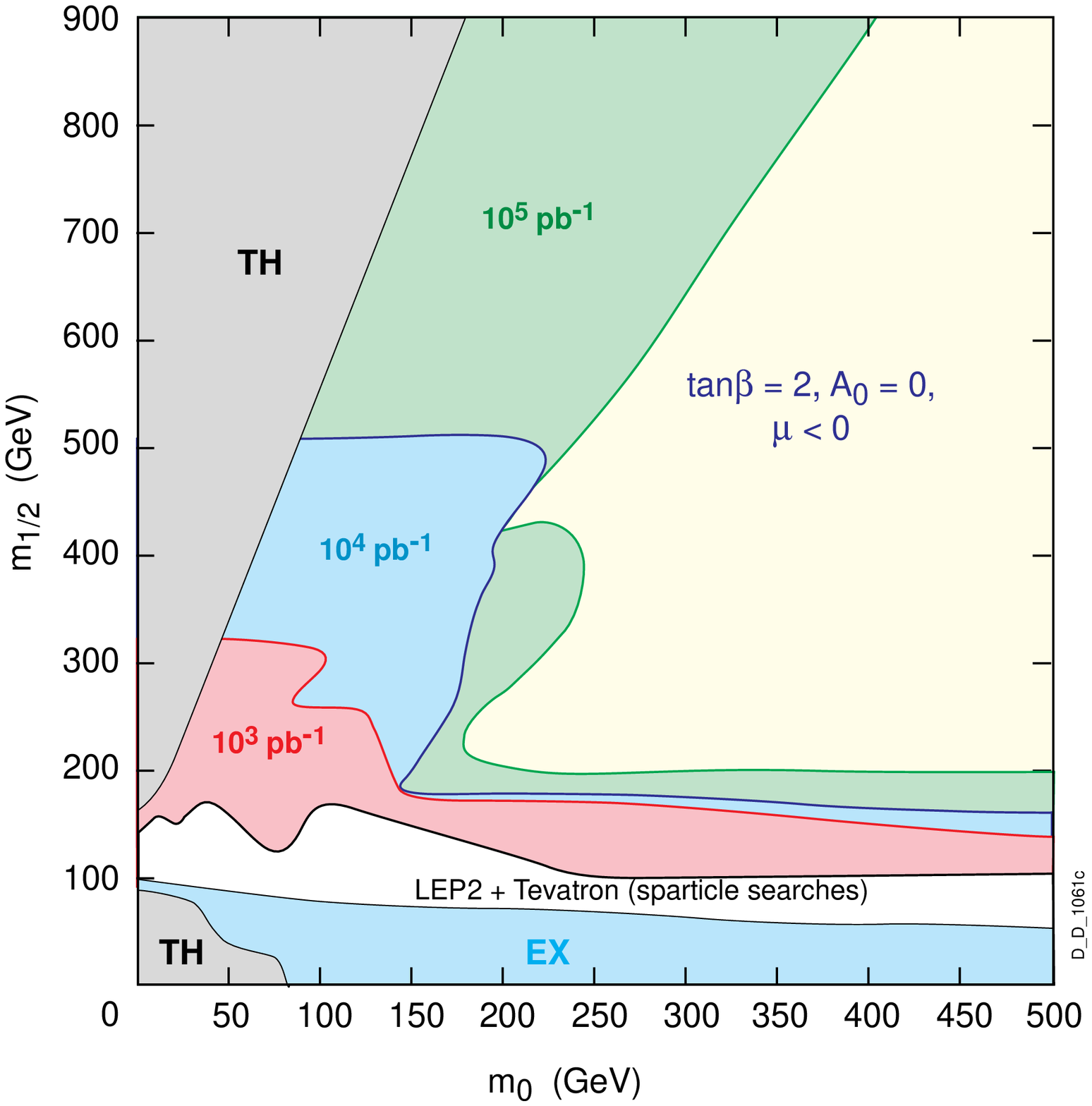,width=8cm}}
\end{center}
 \captive{\it Region of the MSSM parameter space in which
one can detect distinctive `edge' features in the dilepton spectra
due to cascade decays of sparticles, such as $\tilde q \rightarrow q
+ \chi_2, \chi_2 \rightarrow \chi_1 + \ell^+ \ell^-$~\cite{TDR},
for the indicated integrated LHC luminosities.}
\end{figure}

The stage is now set for the
entry of
the next major actor,
the
first-generation
$e^+e^-$ linear collider. It will boast a very clean experimental
environment and
egalitarian production of new weakly-interacting
sparticles, as discussed here by Zerwas~\cite{DESY-ECFA}. Polarization
will be
a useful analysis tool, and $e\gamma, \gamma\gamma$ and $e^-e^-$ colliders will
come `for free'. In many ways, it will be complementary to the LHC. The
trickiest
issue may be how to fix its maximum energy scale. The location of the $\bar tt$
threshold is known, and the precision electroweak data~\cite{LEPEWWG} 
indicate that the ZH
threshold is probably below 300 GeV. This is also expected on the basis
of calculations of the lightest Higgs mass in the MSSM~\cite{HMSSM}.
However, what is the sparticle threshold
(assuming there is one), and how/when will be able to fix it? Flexibility
in the
linear-collider centre-of-mass energy is surely essential. In addition to the
$\bar tt$ and ZH thresholds, obtaining a sample of $10^9$ polarized $Z$ bosons
would provide a very precise determination of $\sin^2\theta_W$, and the $W$
mass
could be measured very precisely at the
$W^+W^-$ threshold~\cite{DESY-ECFA}. However, a centre-of-mass energy of 2
TeV would be
necessary
to ensure full complementarity to the LHC enabling, e.g., the sparticle
spectrum
in Table 1 to be completed.

The first-generation linear collider will enable detailed studies of the Higgs
boson (or the lightest Higgs boson in the MSSM) to be made. Its mass will be
measured to a few parts in $10^4$, and all its major decay modes will be
measured quite accurately~\cite{Battaglia}. This will enable, e.g., a
Standard-Model Higgs boson
to be distinguished from the lightest MSSM Higgs boson, if the heavier MSSM
Higgs
bosons weigh less than several hundred GeV. Even if the centre-of-mass
energy is
restricted to 1 TeV, most of the weakly-interacting sparticles and Higgs bosons
will still be observed directly, and the many spectroscopic measurements will
permit detailed checks of supersymmetric models~\cite{Blaire}.

It has long been clear to me that physics needs a 1-TeV linear
$e^+e^-$ collider, because of its complementarity to the
LHC~\cite{Jerusalem}. It will be
able to
follow up explorations made with the LHC by making many precision measurements.
As already emphasized, the widest possible energy range is desirable. This
implies that any initial lower-energy phase should be extensible to at
least 1
TeV, and running back in the LEP energy range would also be desirable. For the
rest of this talk, I assume that these physics arguments are sufficiently
strong
that a first-generation 1-TeV linear $e^+e^-$ collider will be built.

Nevertheless, there may still be some items on the theoretical wish-list
after the first-generation linear $e^+e^-$ collider. It
would be desirable to have an  accurate direct measurement of the total Higgs
decay width via $s$-channel production, and its mass could be measured much
more
precisely with a muon collider~\cite{Hfact}, as discussed below.
Completing the sparticle
spectrum may require a centre-of-mass energy of 2 TeV or more, as provided by a
second-generation linear $e^+e^-$ collider~\cite{CLIC} or a higher-energy
muon
collider, and
the latter could also produce heavier MSSM Higgs bosons in the direct channel.
Looking further afield, the first glimpse of the 10 TeV energy range could
be
provided by a future larger hadron collider with $E_{cm} \gappeq$ 100
TeV~\cite{VLHC}.

\section{Options for Future Colliders @ CERN}

In mid-1997, the CERN Director-General at the time, Chris Llewellyn Smith
mandated `$\ldots$ a brief written report, $\ldots$, on
possible future
facilities that might be considered at CERN after the LHC'. This should
`$\ldots$ not [be] a major assessment of long-term possibilities'. I would
phrase it as thinking about thinking (about thinking?). The principal options
considered in our report~\cite{EKR} were (i) a next-generation linear
$e^+e^-$
collider with
$E_{cm} \gappeq$ 2 TeV,  based on CLIC technology, (ii) a $\mu^+\mu^-$
collider, ultimately in the multi-TeV $E_{cm}$ range, but perhaps including a
`demonstrator' Higgs factory, and (iii) a future larger hadron collider
(FLHC),
primarily for $pp$ collisions with $E_{cm} \gappeq$ 100 TeV, but perhaps
including options for an $e^+e^-$ top factory and $ep$ collisions in the same
(large) tunnel \footnote{We considered an $ep$ collider in the LEP tunnel to be
already an estalbished CERN option~\cite{LHC-LEP}, and 
in any case not one to be considered a `flagship' project.}.

Starting with the option that we considered least appetizing for CERN, 
if only from the point of view of geography~\cite{EKR}, it seems
apparent that a luminosity of at least 10$^{35}$ cm$^{-2}$s$^{-1}$ would be
required to reap full benefit from a FLHC, perhaps even $10^{36}$
cm$^{-2}$s$^{-1}$ if
$E_{cm} \sim$ 200 TeV. This would pose very severe radiation problems for the
detectors, but such a machine could provide the opportunity to explore the
decade
of mass between 1 and 10 TeV, which history suggests would be a priority after
the LHC.

The default option for the next major project in CERN's future is
probably CLIC, whose physics was first studied
in~\cite{LaThuile}, where its complementarity to the LHC was stressed.
See, in particular, the contributions by Altarelli (p.36), Froidevaux
(p.61), Pauss and myself (p.80), and the review by Amaldi (p.323)
in~\cite{LaThuile}. A study group is now starting to take a further
look at the simulation of benchmark process for CLIC~\cite{BS}.

The CLIC two-beam high-energy $e^+e^-$ collider scheme was presented here
by Delahaye~\cite{CLIC}, so I do not discuss it in detail.
Parameter sets for $E_{cm}$ = 3 and 5 TeV have been developed, and the
central aim
is a cost-effective, affordable strategy for such a higher-energy linear
collider, since the key CLIC advantages of a high accelerating gradient
and (relatively) simple
components are not needed for a first-generation $E_{cm} \lappeq$ 1 TeV linear
collider. Two CLIC test facilities have already been built and operated
successfully, CTF1 and CTF2~\cite{CLIC}. However, the need for at least
two more demonstrator projects is foreseen
before construction of CLIC itself can be envisaged. These are CTF3 in the
years
2000 to 2005, to demonstrate the acceleration potential in a 0.5 GeV
machine, and
then CLIC1 in the years 2005 to 2009, which should attain 75
GeV~\cite{CLIC}. Recall also that no major
capital investment money will become available at CERN before 2009, because
of the
LHC payment schedule. For both the reasons in the two previous sentences,
CLIC is
necessarily on a longer time scale than that proposed for first-generation
linear
collider projects such as TESLA, the JLC or the NLC.

A CERN geological study has indicated that the tunnel for a linear collider
$\sim$ 30 km long could be excavated parallel to the Jura, entirely in suitable
molasse rock: similar conclusions were reached in a study conducted for
Swissmetro (the group that proposes to build a high-speed underground railway
connecting Geneva and other major Swiss cities)~\cite{EKR}. Also, even a
$E_{cm}$ = 4
TeV $\
\mu^+\mu^-$  collider would fit comfortably within the area bounded by the
existing SPS and LEP/LHC tunnels. On the other hand, it is difficult to see how
even a high-field FLHC with $E_{cm}$ = 100 TeV (which would require a
tunnel
circumference in excess of 100 km) could be accommodated in the
neighbourhood of
CERN.

As far as technological maturity is concerned, even though several hurdles need
to be crossed before the CLIC technology is mature -- for example, the beam
delivery system has hardly been studied -- it may be the closest to mass
shell of
the next-generation collider concepts. The technology required for a FLHC
exists in
principle, but the key problem is to reduce the cost per TeV by an order of
magnitude compared to the LHC. This will require innovative ideas for
tunnelling,
as well as magnets and other machine components~\cite{VLHC}.

The most speculative option we considered was a $\mu^+\mu^-$ collider, many of
whose components are at best extrapolations of current technologies, with many
others not existing in any form. Considerable R\&D is required even to
establish
the plausibility of the
$\mu^+\mu^-$ collider concept.
This challenge spurred the formation some years ago in the US of the Muon
Collider Collaboration~\cite{MCC}, which groups a hundred or more
physicists and engineers
and has proposed R\&D projects, notably on ionization
cooling~\cite{MUCOOL}. Until recently,
there was little activity in Europe on muon colliders, although some individual
CERN staff members worked with the Muon Collider Collaboration. This disparity
led RECFA to commission in 1998 a prospective study of $\mu^+\mu^-$ colliders,
whose brief was to specify the physics case, to identify areas requiring
R\&D, and
look for potential European resources outside CERN and DESY.

The corresponding report~\cite{MCYB} produced in early 1999 proposed a
three-step scenario
for physics with  muon storage rings at CERN, illustrated in Fig. 6. The first
step would be a
$\nu$ factory~\cite{nufact}, in which an intense proton source would be
used to produce
muons,
that would be captured and then cooled by a limited factor, before being
accelerated and stored in a ring and allowed to decay, without being brought
into collision. Such a $\nu$ factory had not been considered
in~\cite{EKR}: the physics interest in such a machine had been amplified
in the mean time, in particular by the emerging evidence for atmospheric
neutrino oscillations. The big advantages over a
conventional $\nu$ beam produced directly by hadronic decays are that the
$\nu$ beams produced by $\mu$ decay would have known fluxes, flavours, charges
and energy spectra, and would  comprise equal numbers of $\nu_\mu$ and
$\bar\nu_e$ (or $\bar\nu_\mu$ and $\nu_e$). Such a $\nu$ factory would
surely
be the `ultimate weapon' for $\nu$ oscillation studies. This could be
followed
by a second step (or steps), namely a Higgs factory (or
factories)~\cite{Hfact}, which
could
measure accurately the mass, width and other properties of a Standard Model
Higgs via its direct $s$-channel production, and thus distinguish between it
and the lightest Higgs in the MSSM, strongly constraining its parameter
space in
the latter case. A second factory operating on the adjacent peaks of the other
neutral $H$ and $A$ Higgs bosons of the MSSM would also be interesting,
possibly
opening a novel window on CP violation in the Higgs sector. The third step
would
be a high-energy frontier $\mu^+\mu^-$ collider. Its advantages over an
$e^+e^-$
collider would include superior beam-energy resolution and
calibration~\cite{MCYB}, whereas
an $e^+e^-$ collider such as CLIC would also offer beam polarization and the
possibilities of
$e\gamma, e^-e^-$ and $\gamma\gamma$ collisions.

The $\nu$ and Higgs factories are discussed in the next two sections of this
talk, followed by a comparison of the $e^+e^-$ and $\mu^+\mu^-$ strategies for
attaining the high-energy frontier in lepton-lepton collisions.
\begin{figure}
\begin{center}
\mbox{\epsfig{file=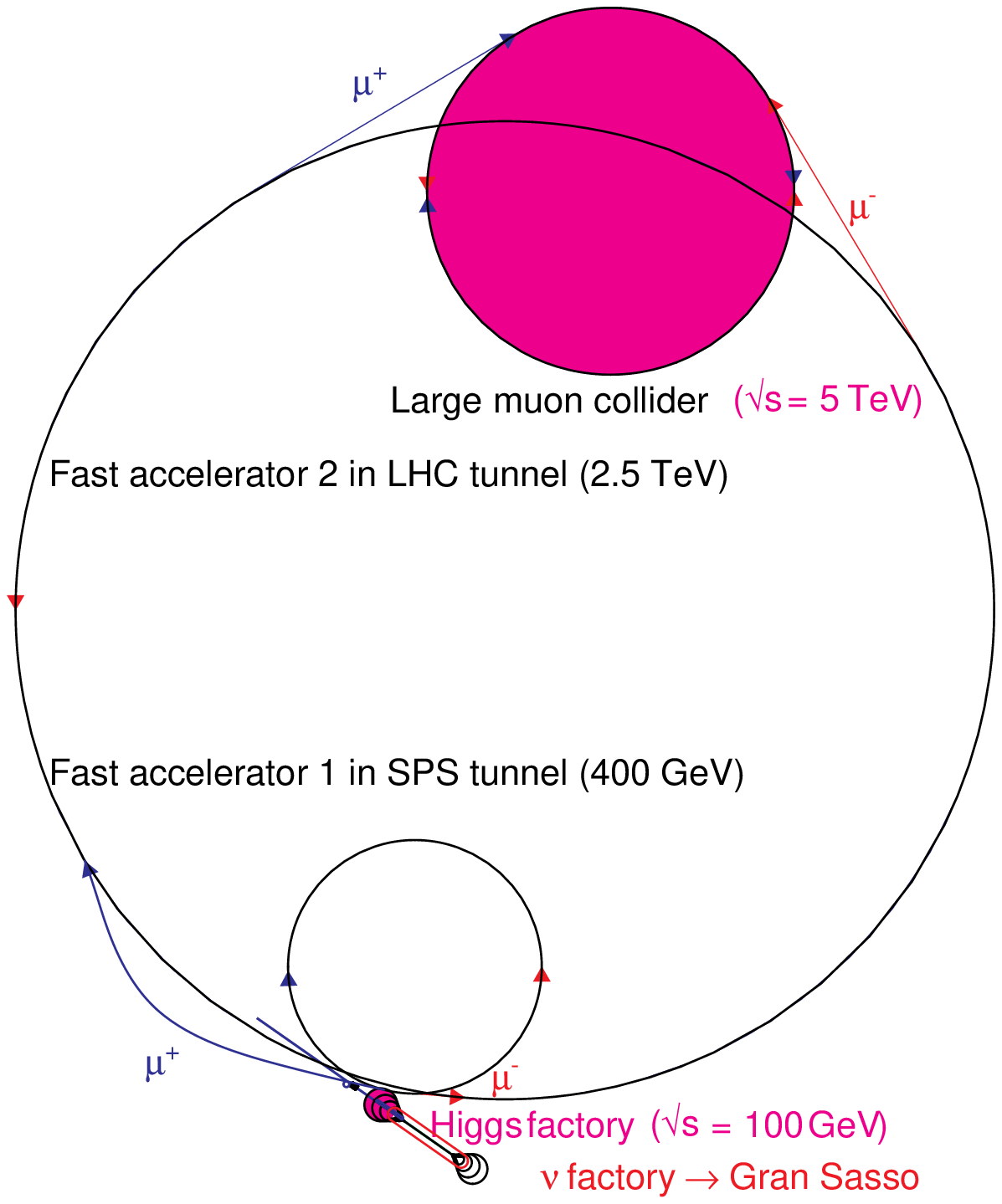,width=8cm}}
\end{center}
 \captive{\it Schematic layout of a possible three-step neutrino
storage ring complex at CERN, including a $\nu$ factory,
a Higgs factory and a possible high-energy frontier muon
collider~\cite{MCYB}.}
\end{figure}

\section{A Neutrino Factory}

The basic concept of a $\nu$ factory~\cite{MCYB} starts with an intense
low-energy proton
driver providing one to 20 MW (say 4 MW) of beam power, based either on a
linac or
a rapid-cycling synchrotron. Beam energies between a few and 30 GeV are
discussed
actively~\cite{Lyon}, and the 50~GeV JHF project could be an interesting
prototype project~\cite{JHF}. A typical source intensity would be
1.5$\times 10^{15} p/s$ at 16
GeV.
This beam is used primarily to produce pions, which are allowed to decay into
muons that must
 be captured, and a typical rate might be 0.2 $\mu/p$. Of these, about a half,
namely 0.1 $\mu/p$, would survive being cooled by a factor 10 to 100 in phase
space. These would then be accelerated, perhaps by a recirculating linac,
to the
chosen storage energy, which would probably lie in the range 10 to 50 GeV. As
seen in Fig. 7, the storage `ring' itself could be quite irregular, with
two or
three straight sections that would each yield ${\cal O}(3\times 10^{20})
\bar\nu_\mu$ and
$\nu_e$ per year (or
$\nu_\mu$ and $\bar\nu_e$) in any given direction. Two of these directions
should
be those of large underground detectors located far away, perhaps one at
several
hundred kilometres and one at several thousand, optimized for oscillation
studies. There could be a third detector close to the `ring', optimized
for
Standard Model studies with neutrinos~\cite{King}.

\begin{figure}
\begin{center}
\mbox{\epsfig{file=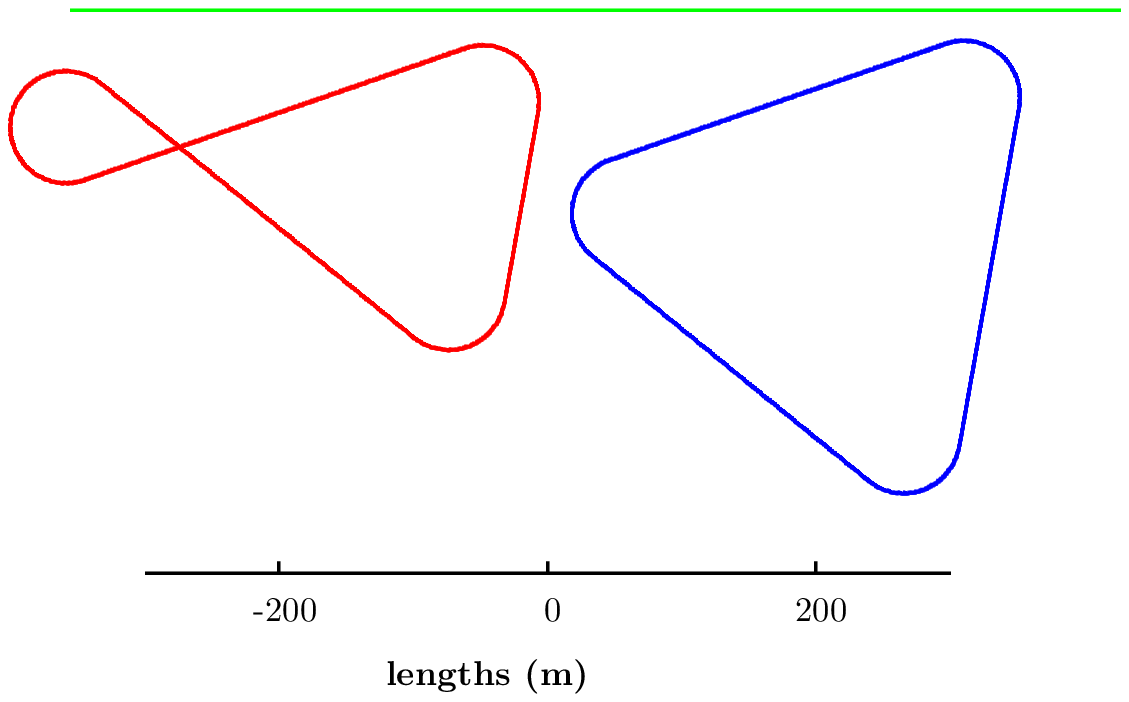,width=10cm}}
\end{center}
\captive{\it Schematic geometries of possible muon storage
`rings' for a $\nu$ factory~\cite{NSF}, with straight sections
pointing towards long-, very-long-, and short-baseline experiments.
The upper horizontal line is the ground surface.}
\end{figure}

There are many accelerator
issues for
such a
concept~\cite{NSF}. In
addition to the choice
of a rapid-cycling (how many Hz are possible?) synchrotron or a linac
(could
it re-use the LEP superconducting RF?), these include: the target -- a liquid
Mercury jet has been studied, and solid metal wires or strips have been
proposed,
pion capture -- which would require a 20 Tesla solenoid, a monochromator --
which
would require high-field (pulsed?) RF working in a high-radiation environment
with a strong magnetic field, a cooling channel -- which need only compress the
muon phase space by a factor of 10 to 100 rather than the $10^6$ needed for a
collider, and a recirculating linac to accelerate the muons -- which might be
another application for the LEP RF.

Before discussing the primary physics objective of neutrino oscillations, we
first review basic formulae for neutrino mixing~\cite{NKS}. Analogously to
the Cabibbo-Kobayashi-Maskawa mxing of quarks, one has
\beq
\left(\matrix{\nu_e\cr\nu_\mu\cr\nu_\tau}\right) =
\left(\matrix{c_{12} c_{13}\phantom{xxxxxxxxx} &  c_{13}
s_{12}\phantom{xxxxxxxxxx} &  s_{13}
\cr -c_{23} s_{12} e^{i\delta} - c_{12} s_{13} s_{23} & c_{12} c_{23}
e^{i\delta}
-s_{12} s_{13} s_{23} & c_{13} s_{23} \cr
s_{23} s_{12} e^{i\delta}  - c_{12} c_{23} s_{13} & -c_{12} s_{23}
e^{i\delta}  -c_{23}
s_{12} s_{13} & c_{13} c_{23} }\right)
~~\left(
\matrix{\nu_1\cr\nu_2\cr\nu_3}\right)
\label{one}
\eeq
where the $\nu_i: i = 1,2,3$ are mass eigenstates. In addition to the
CP-violating phase $\delta$ in (\ref{one}), there are also two CP-violating
relative Majorana phases that are unobservable at energies $E \gg m_{\nu_i}$,
but need to be taken into account in considering the constraints imposed by
double-$\beta$ decay. In the simplified limit: $\Delta m^2_{12} \ll E/L \sim
\Delta m^2_{23}$, one has the following oscillation
probabilities~\cite{DGH}:
$$
P(\nu_e\rightarrow\nu_\mu) = \sin^2\theta_{23} \sin^2 \theta_{13} \sin^2
\left({\Delta m^2_{23} L\over 4E}\right)
\eqno{(2a)}
$$
$$
P(\nu_e\rightarrow\nu_\tau) = \sin^2\theta_{23} \sin^2 \theta_{13} \sin^2
\left({\Delta m^2_{23} L\over 4E}\right)
\eqno{(2b)}
$$
$$
P(\nu_\mu\rightarrow\nu_\tau) = \sin^4\theta_{13} \sin^2 \theta_{23} \sin^2
\left({\Delta m^2_{23} L\over 4E}\right)
\eqno{(2c)}
$$
\addtocounter{equation}{1}
and the same for the time-reversed transitions $P(\nu_\mu\rightarrow\nu_e)$,
etc., since the CP-violating phase $\delta$ is observable only when both
$\Delta
m^2_{12}$ and $\Delta m^2_{23}$ are large.

Figure 8 shows the sensitivity in the $(\sin^2\theta_{23}, \Delta
m^2_{23})$ plane of a neutrino factory, for 
oscillations in the atmospheric range
in a `long'-baseline experiment: $L$ = 730 km, in both the $\nu_\mu$ 
disappearance and appearance modes for different values of
$\theta_{13}$~\cite{DGH}. The
updated region allowed by Super-Kamiokande $(\Delta m^2_{23} \gappeq 2\times
10^{-3}$ eV$^2$) is comfortably within reach. Figure 9 shows the sensitivity in
the $(\sin^2\theta_{13},\Delta m^2_{23})$ plane for $\nu_\mu
\leftrightarrow\nu_e$ oscillations, again in both appearance and disappearance
modes~\cite{DGH}. For the current range of $\Delta m^2_{23}$, the
appearance mode has
access
to values of $\sin^2\theta_{13}$ for below the current upper limits imposed by
Super-Kamiokande and Chooz, and also far below what can be reached with a
conventional hadronic-decay $\nu_\mu$ beam, which contains $\nu_e$
contamination
close to the 1 \% level.

\begin{figure}
\begin{center}
\mbox{\epsfig{file=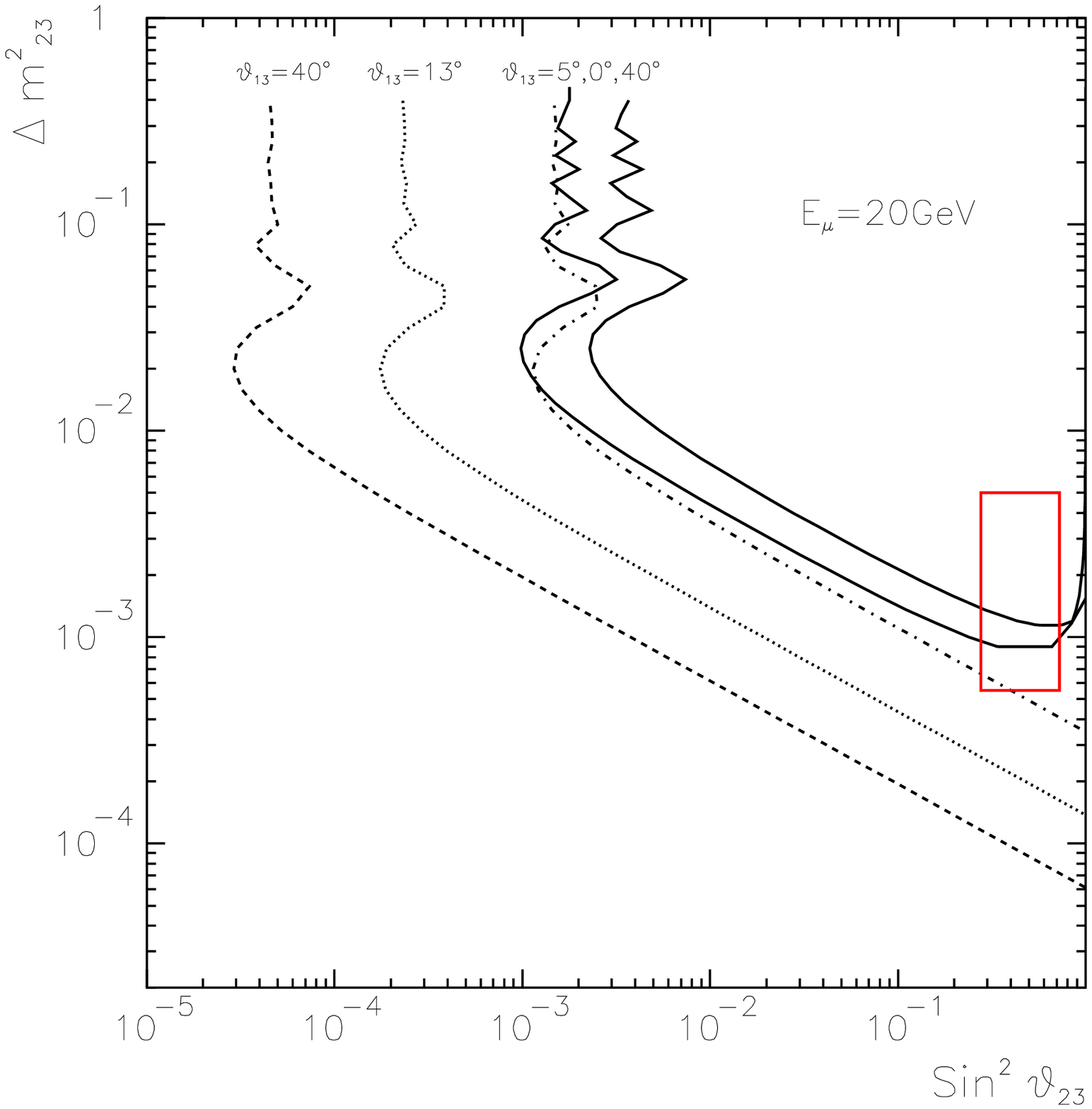,width=8cm}}
\end{center}
 \captive{\it Sensitivity in the $(sin^2 \theta_{23}, \Delta
m^2_{23})$ plane of a long-baseline $\nu$ factory detector
looking in a $\nu_\mu$ beam for $\nu_\mu$ disappearance (solid line) or
$\nu_\mu$
appearance (dashed line)~\cite{DGH}, for different values of
$\theta_{13}$.}
\end{figure}

\begin{figure}
\begin{center}
\mbox{\epsfig{file=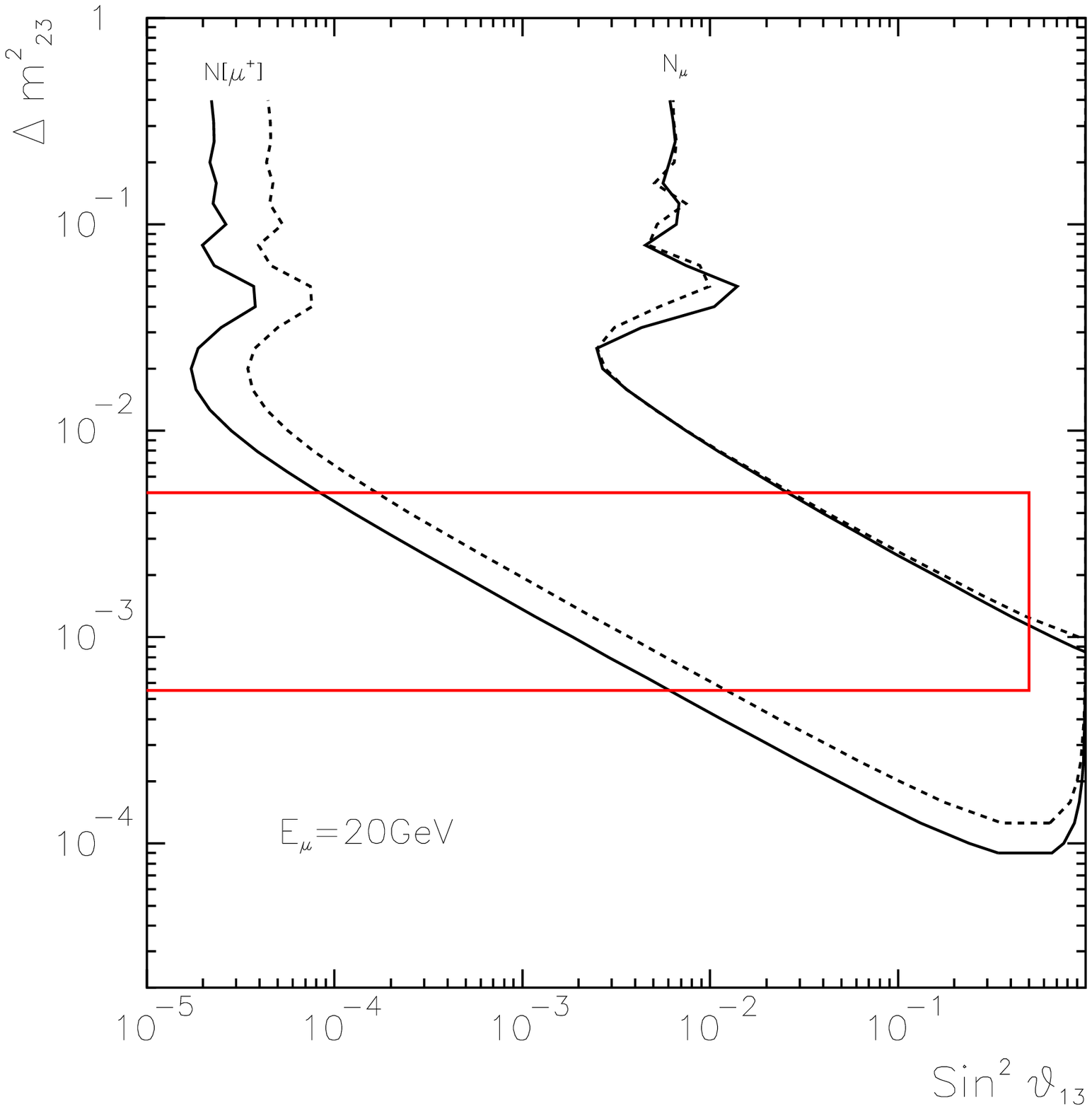,width=8cm}}
\end{center}
\captive{\it Sensitivity in the $(sin^2 \theta_{13}, \Delta
m^2_{23})$ plane of a long-baseline $\nu$ factory detector
looking in a $\nu_\mu$ beam for $\nu_\mu$ disappearance and $\nu_\mu$
appearance (more
sensitive) for $\theta_{23} = \pi/4, \pi/6$ (dashed, solid lines)~\cite{DGH}.}
\end{figure}

The fluxes obtained from a
$\nu$
factory are so intense that
experiments over a
range of several thousand kilometres become feasible~\cite{BCR}, and new
domains of
oscillation phenomena become accessible, perhaps including the CP- and
T-violating
asymmetries~\cite{CPandT}:
\beq
A_{CP} \equiv
{P(\nu_\mu\rightarrow\nu_e) -
P(\bar\nu_\mu\rightarrow\bar\nu_e)\over
P(\nu_\mu\rightarrow\nu_e) +
P(\bar\nu_\mu\rightarrow\bar\nu_e)}~,
\quad\quad A_T\equiv
{P(\nu_\mu\rightarrow\nu_e) -
P(\nu_e\rightarrow\nu_\mu) \over
P(\nu_\mu\rightarrow\nu_e) +
P(\nu_e\rightarrow\nu_\mu)}
\label{three}
\eeq
The CP-violating asymmetry may be large if both $\Delta m^2_{12}$ and
$\theta_{12}$ are large:
\beq
A_{CP} \simeq
{4\sin\theta_{12}\sin\delta\over\sin\theta_{13}}~~\sin~~\left({\Delta
m^2_{12} L
\over 2E}\right)
\label{four}
\eeq
and may be observable if the solar-neutrino deficit is due to the
large-mixing-angle MSW solution. On the other hand, measuring $A_T$ would
require
$e^\pm$ discrimination, which is difficult in a muli-kiloton detector.
Measurements of
$A_{CP}$ must contend with the fact that the Earth is not  CP-invariant, so
that
matter effects also contribute to $A_{CP}$, e.g.,
\beq
A_{CP}^{MSW} \simeq 0.7 \times 10^{-6} \times {L^2(km^2)\over E({\rm GeV})}
\label{five}
\eeq
for $\Delta m^2_{23} = 3\times 10^{-3}$ eV$^2$ and
$\Delta m^2_{12} = 3\times 10^{-4}$ eV$^2$. This means that the matter effect
(\ref{five}) dominates for $L \gappeq$ 4000 km, as seen in Fig.
10~\cite{Donini}.

\begin{figure}
\begin{center}
\mbox{\epsfig{file=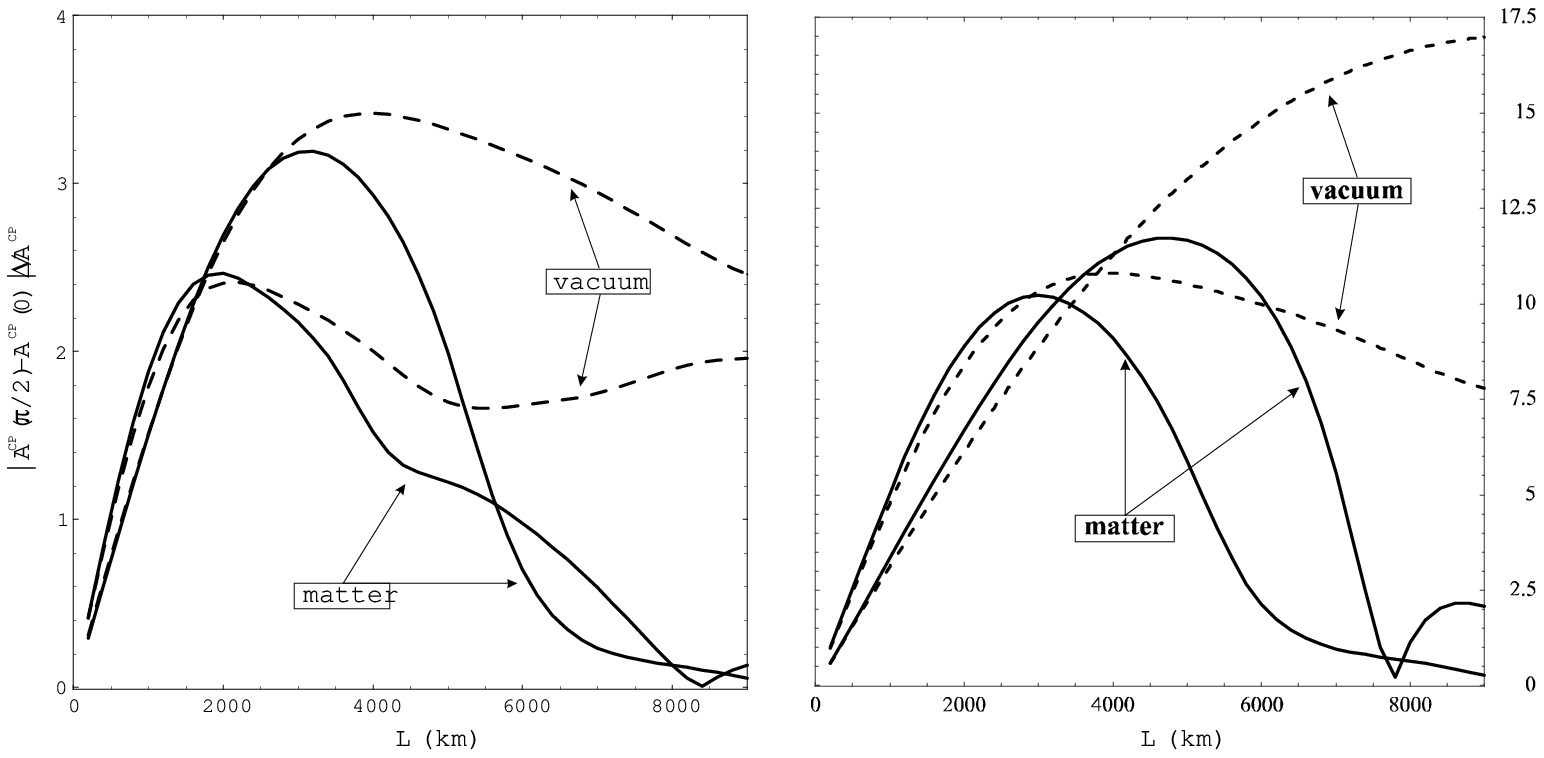,width=14.5cm}}
\end{center}
\captive{\it Significance of the observation of a CP-violating 
asymmetry $A_{CP}$ with $2 \times 10^{20}$ (left panel) or $2 \times
10^{21}$ neutrinos (right panel), including (solid lines) or discarding
(dashed
lines) matter effects, for the mixing parameters described
in~\cite{Donini}.} 
\end{figure}

The preferred baseline for observing the intrinsic $A_{CP}$ (\ref{four})
seems
to be $L \sim$ 2000 to 3000 km, and it seems that a $5-\sigma$ effect
could be
observed throughout the large-angle MSW region with a 50 kt detector
operated for
five years in conjunction with a 20 MW source.
Further studies will be needed to optimize the choice of $L$, depending,
e.g., on
what we learn from future solar-neutrino experiments,
KamLAND~\cite{KamLAND} and possibly atmospheric neutrinos
in a low-threshold detector such as ICANOE~\cite{ICANOE}.
It could well be that
CP violation is unobservable, e.g., if the deficit is due to either the
small-angle MSW solution or vacuum oscillations, or if $\delta$ is small. The
matter oscillations that dominate at larger
$L$ in Fig. 10 might be of supplementary interest, either in their own
right, to
fix the sign of
$\Delta m^2_{23}$ or even to probe the internal structure of the Earth.

Other interesting particle physics~\cite{AIP} would also be possible with
the intense
proton
driver needed for a $\nu$ factory. For example, it might be possible to improve
by several orders of magnitude the current upper limits on
charged-lepton-flavour
violation in the processes $\mu\rightarrow e\gamma$, $\mu\rightarrow 3e$ and
$\mu Z\rightarrow eZ$. Such experiments could explore the range of
interest to
supersymmetric GUT models of $\nu$ oscillations, as seen in Fig.
11~\cite{EGLLN}. Other
quantities of interst for possible physics beyond the Standard Model
include the
muon's anomalous magnetic moment -- can an experiment with better sensitivity
than that presently running at BNL be envisaged? and can the uncertainties
in the
hadronic contributions be controlled sufficiently to look, e.g., for possible
supersymmetric contributions? It might also be   possible to look for an
electric
dipole moment -- measuring it with a precision $\sim 10^{-24}$ e.cm would have
significance comparable to the present limit of $4\times 10^{-27}$ e.cm for the
electron, since in many models
$d_\mu/d_e \simeq m_\mu/m_e \simeq$ 200. Other physics opportunities might be
offered by rare $K$ decays, such as $K^0_L\rightarrow\pi^0 \bar\nu\nu,
K^+\rightarrow\pi^+\bar\nu\nu$, $K \rightarrow \pi \ell^+\ell^-$, $K\rightarrow
\mu e$ and $K\rightarrow\pi\mu e$, if the beam energy of the proton driver is
sufficiently high to produce kaons copiously.

\begin{figure}
\begin{center}
\mbox{\epsfig{file=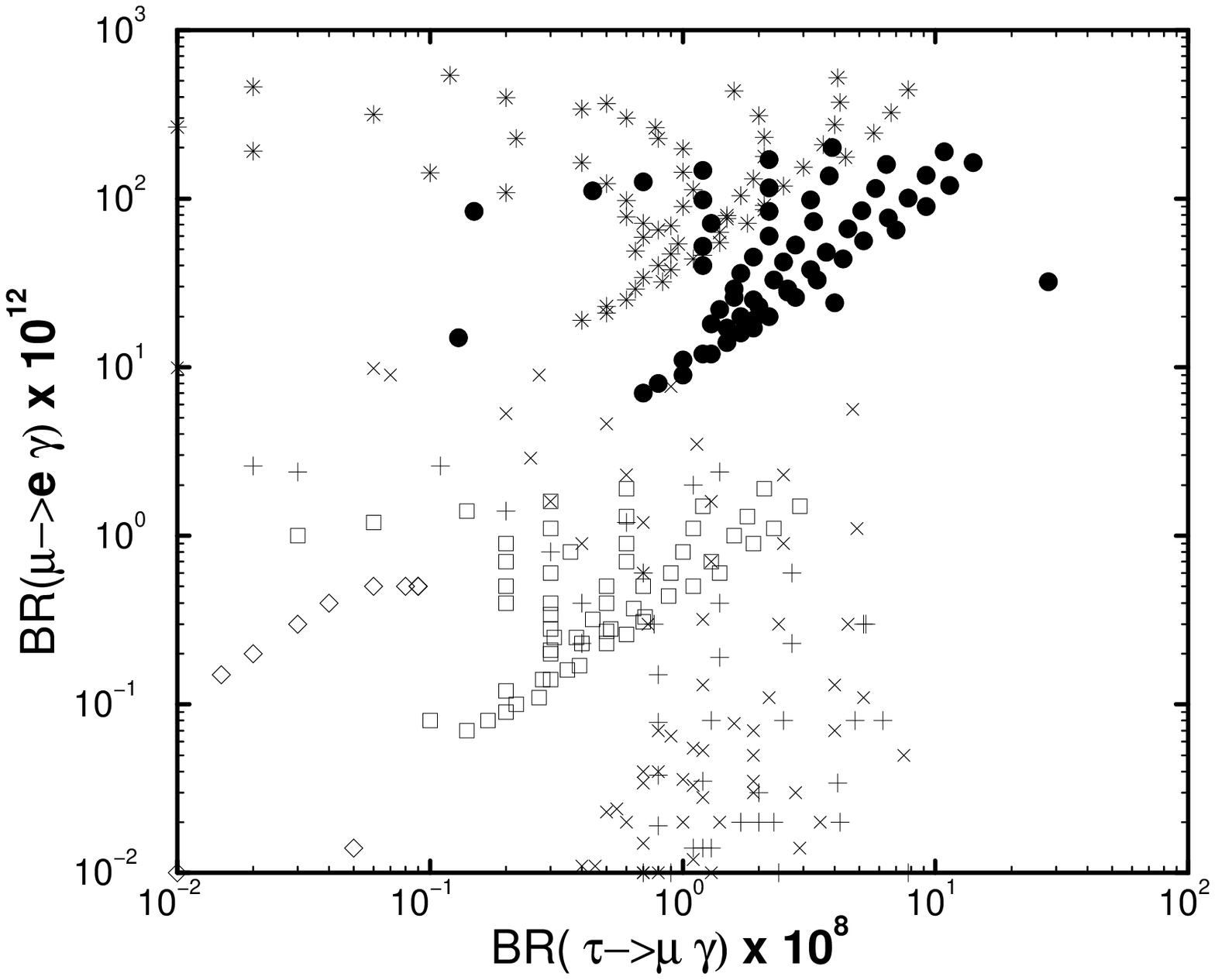,width=8cm}}
\end{center}
\captive{\it Rates for $\mu \rightarrow e \gamma$ and
$\tau \rightarrow \mu \gamma$ decay in some generic
supersymmetric GUT models inspired by the Super-Kamiokande
data on neutrino oscillations, showing opportunities both
for intense $\mu$ beams and for intense $\tau$ sources, such as the
LHC~\cite{EGLLN}.}
\end{figure}

As
has
already been mentioned,
`traditional' neutrino
physics could be pursued
using a nearby detector, with very high statistics~\cite{King}. This
could be of
interest for
measuring $\sin^2\theta_W$ and Cabibbo-Kobayashi-Maskawa matrix elements
very
precisely, and one could perhaps even use a polarized target and probe the
nucleon spin in a novel way. One could also perform $\mu$ scattering
experiments
with a target in the
$\mu$ `ring' itself. How much of the ELFE physics programme~\cite{ELFE} 
could be
addressed by
these NULFE and MULFE options?

Beyond particle physics, the proton driver could be used for many other
experiments of interest to nuclear physicists, e.g., on muonic atoms, $\mu$
capture and radioactive beams~\cite{AIP}. Also the necessary
high-intensity proton-beam
technology would have much in common with requirements for other classes of
applications, e.g., an advanced Spallation Neutron Source, radioactive
waste
disposal and the concept of an energy amplifier.

Therefore, this first step in the scenario for physics with muon storage rings
could be of interest to a broad community.

\section{Higgs Factories}

The second step -- the Higgs factory (or factories) requires much more beam
cooling, and relies on the relatively large $\mu^+\mu^-H$ coupling:
$\Gamma(H\rightarrow\mu^+\mu^-) \sim 4\times 10^4\Gamma(H\rightarrow e^+e^-)$,
and the superior beam-energy calibration and small energy spread to render
direct
$s$-channel Higgs production $\mu^+\mu^-\rightarrow H\rightarrow X$ measurable.
Neglecting beam energy spread,  the Higgs line shape would
be~\cite{Hfact,MCYB}
\beq
\sigma_H(s) \simeq {4\pi \Gamma(H\rightarrow\mu^+\mu^-)\Gamma(H\rightarrow
X)\over (s-m^2_H)^2 + M^2_H \Gamma^2_H}
\label{six}
\eeq
Since the natural width $\Gamma_H$ of the Higgs is measured in MeV if $m_H
\simeq$ 100 GeV, a beam-energy spread $\lappeq 10^{-4}$ is desirable.
Fig. 12 shows what could be attainable using a Higgs factory for a
Standard
Model Higgs weighing 110 GeV~\cite{MCYB}. Thanks to the natural
beam-energy calibration
provided by the decay of polarized muons, it will be possible to calibrate the
beam energy with a precision of 5 keV each fill~\cite{mupol}, enabling
$m_H$ to be measured
with an error of 0.1 MeV, and $\Gamma_H$ could be measured with an error of 0.5
MeV by making a three-point scan~\cite{MCYB}.

\begin{figure}
\begin{center}
\mbox{\epsfig{file=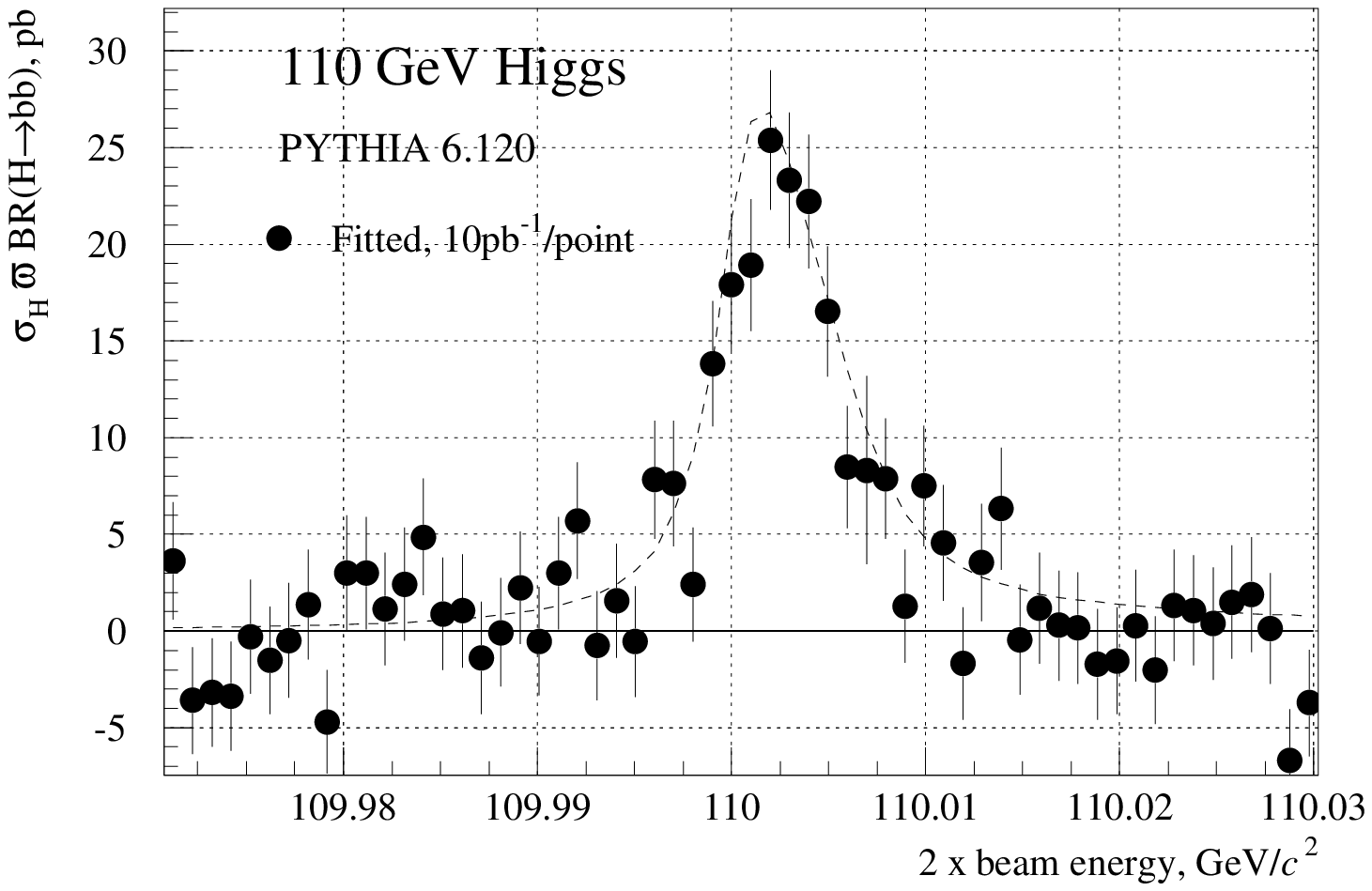,width=8cm}}
\end{center}
\captive{\it Simulated measurements of the direct-channel line
shape at a Higgs factory, for a Higgs boson with
mass 110~GeV, assuming 10~pb$^{-1}$ of integrated luminosity but excluding
the possible beam energy spread~\cite{MCYB}.}
\end{figure}

This should
enable the nature of
the Higgs boson to be
clarified. For example,
the lightest MSSM Higgs boson typically has a much larger width than a Standard
Model Higgs boson of the same mass, particularly for small $m_A$. As seen in
Fig. 13, the parameters of the MSSM could be inferred~\cite{MCYB} much
more precisely from
$\mu^+\mu^-$ measurements than with the LHC and/or an $e^+e^-$ linear collider.

\begin{figure}
\begin{center}
\mbox{\epsfig{file=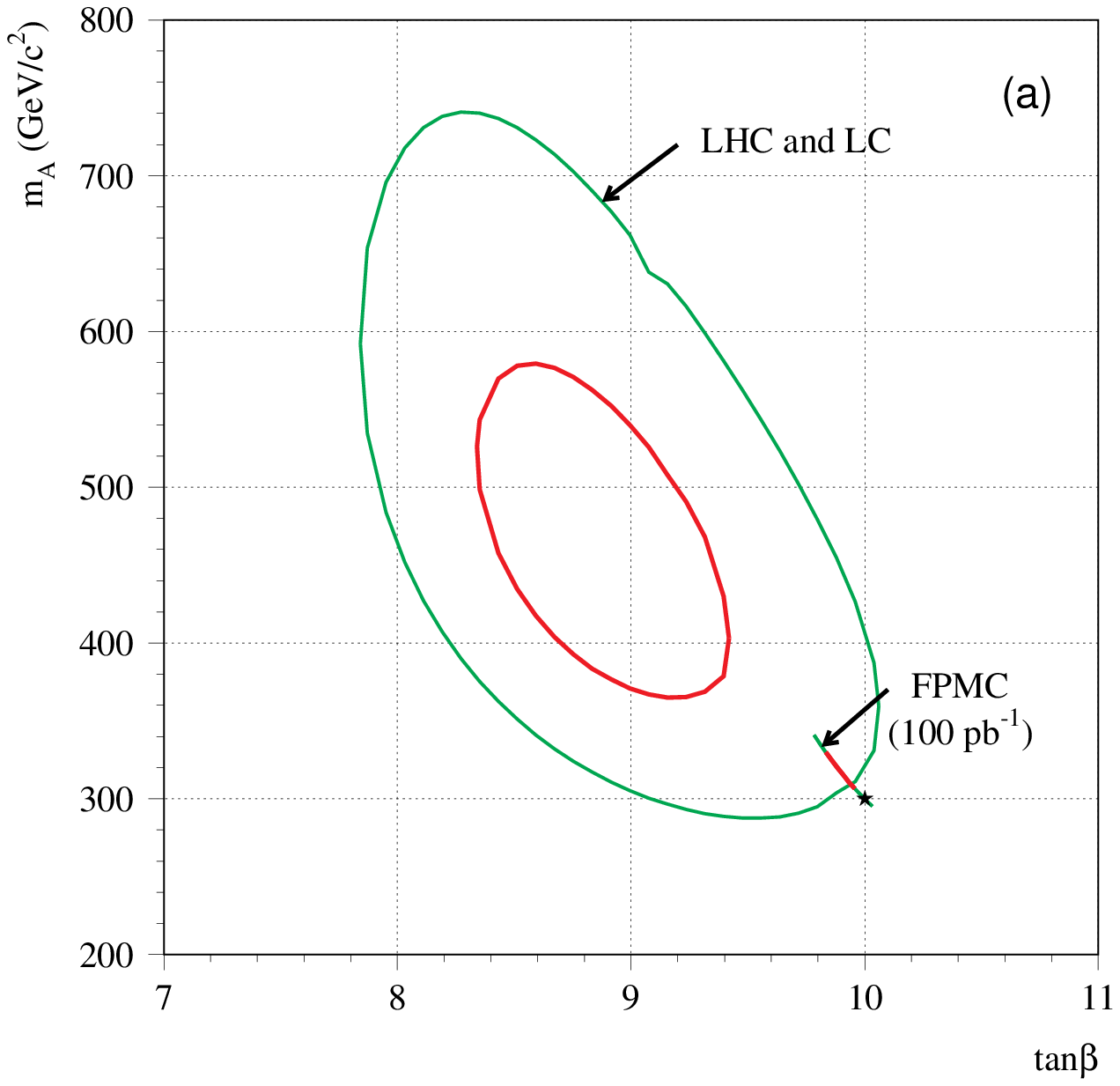,width=8cm}}
\end{center}
\captive{\it Estimated precision with which
direct-channel measurements of the Higgs line
shape at a Higgs factory could be used to constrain
MSSM parameters, compared with analogous estimates for the
LHC and an $e^+ e^-$ linear collider~\cite{MCYB}.}
\end{figure}

A
second
Higgs factory could then
be constructed to explore
the twin peaks
of the
$H$ and $A$ in the MSSM. Fig. 14 shows the result of a case study based on
a
coarse scan of $\pm$ 60 GeV in a previously-established mass region with $1
pb^{-1}$/GeV, followed by a fine scan of six points with $25$ 
pb$^{-1}$/point~\cite{MCYB}.
These would be sufficient to establish the peak cross sections with precisions
$\Delta \sigma^{H,A}_{peak}/\sigma^{H,A}_{peak} = \pm 1 \%$, $\Delta m_{H,A} =
\pm$ 10 MeV, $\Delta\Gamma_{H,A} = \pm$ 50 MeV. Many detailed studies of
$H$ and
$A$ decay modes would be possible, and one of the enticing possibilities
would be
to look for CP violation in their decays~\cite{Pilaftsis}.

\begin{figure}
\begin{center}
\mbox{\epsfig{file=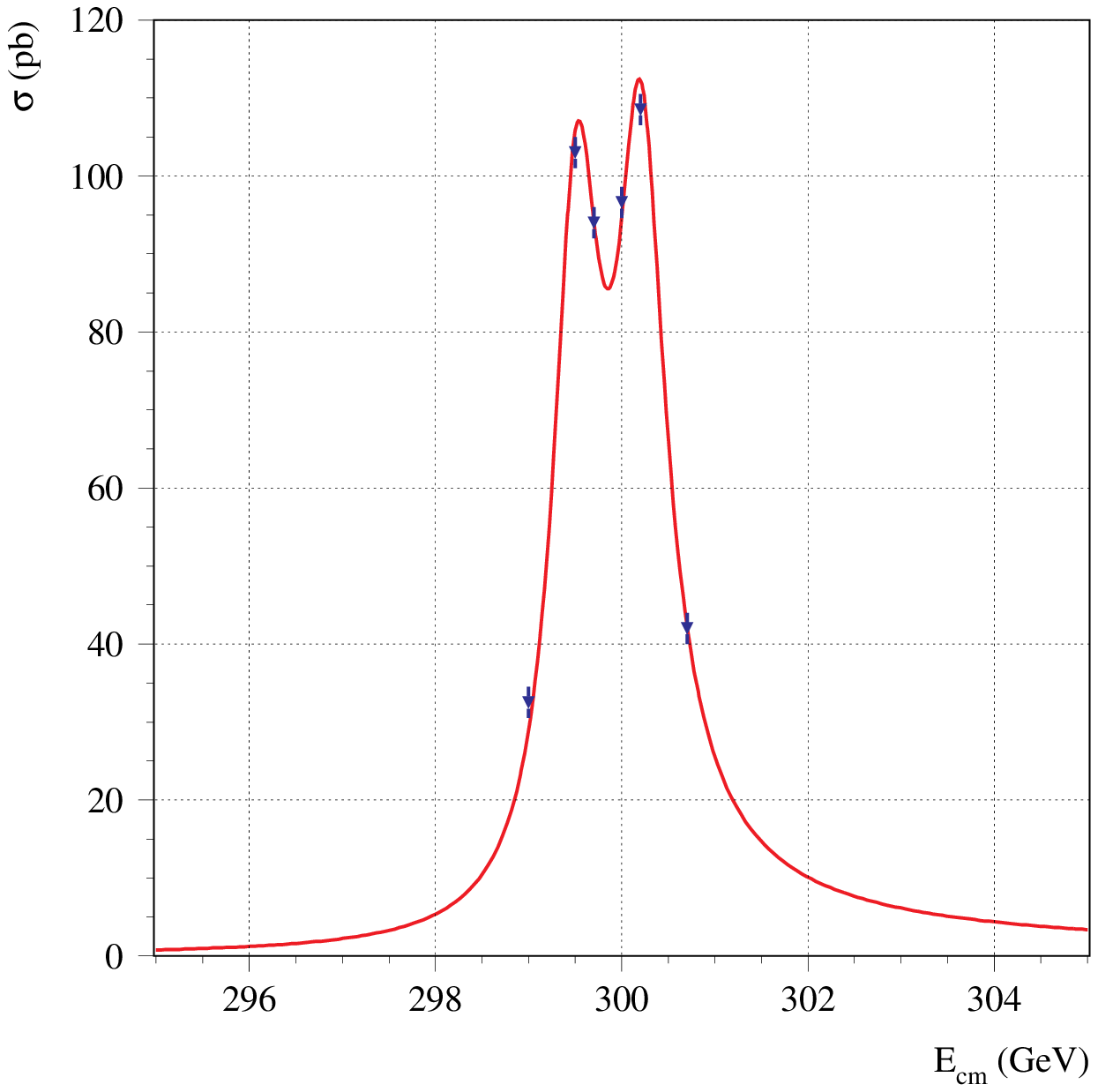,width=8cm}}
\end{center}
\captive{\it Simulated measurements of the direct-channel
production of the heavier MSSM Higgs bosons $(h, A)$ at a Higgs
factory, assuming $m_A = 300$~GeV, tan$\beta = 10$, 
25~pb$^{-1}$ of integrated luminosity per point,
and a beam-energy spread
of $3 \times 10^{-5}$~\cite{MCYB}.}
\end{figure}

\section{The High-Energy
Frontier}

A lepton collider with several TeV of centre-of-mass energy would have a
physics
reach extending beyond the LHC in many respects. What might be interesting
physics at that time, and how would high-energy $e^+e^-$ (e.g., CLIC) and
$\mu^+\mu^-$ colliders compare? Their effective mass reaches may be assumed
to be
similar: $E_{cm}$ for CLIC might be limited for both financial and technical
reasons, and $E_{cm}$ for a $\mu^+\mu^-$ collider might be limited by the
danger
of neutrino radiation~\cite{nurad}, as discussed later.

The difference between the lepton flavours might play a role in some physics
processes, for example in the context of $R$-violating
supersymmetry~\cite{MCYB}, where
$e^+e^-$ and $\mu^+\mu^-$ colliders are sensitive to different couplings.
We have
already seen how the larger Higgs-$\mu^+\mu^-$ coupling could confer advantages
on a lower-energy $\mu^+\mu^-$ collider. The same would be true of a
higher-energy $\mu^+\mu^-$ collider if, for example, $m_A$ were very large,
i.e.,
above 2 TeV. The smaller energy spread and better energy calibration of  a
higher-energy $\mu^+\mu^-$ collider could also be interesting, for example for
threshold measurements. One example studied~\cite{MCYB} was the reaction
$\mu^+\mu^-\rightarrow\chi^+\chi^-$,  where the threshold cross section is much
more sensitive to $m_{\chi^\pm}$ than is $e^+e^- \rightarrow 
\chi^+\chi^-$:
\beq
{d\sigma\over dm_{\chi^\pm}} \bigg\vert_{\mu^+\mu^-} \simeq 24 \times
{d\sigma\over dm_{\chi^\pm}} \bigg\vert_{e^+e^-}
\label{seven}
\eeq
after inclusion of initial-state radiation and beamstrahlung effects, as
shown in
Fig. 15. There could also be some advantage in the study of narrow
resonances, as
might occur in some models of strongly-coupled Higgs sectors and/or extra
dimensions~\cite{MCYB}.

\begin{figure}
\begin{center}
\mbox{\epsfig{file=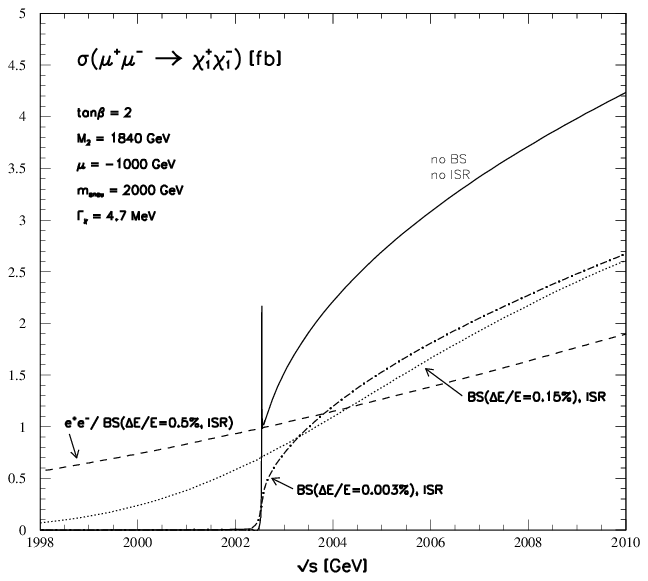,width=8cm}}
\end{center}
\captive{\it Calculation of the $\mu^+ \mu^- \rightarrow \chi^+
\chi^-$ chargino-pair threshold, showing the effects of
initial-state radiation (ISR) and beamstrahlung (BS)~\cite{MCYB}.}
\end{figure}

On the
other hand, there
are some instances where the
availability of
$e\gamma$,
$\gamma\gamma$ and $e^-e^-$ collisions with an $e^+e^-$  collider could be
advantageous. Table~2~\cite{MCYB} lists some relevant physics topics,
summarizes the
principal capabilities of high-energy $\mu^+\mu^-$ and $e^+e^-$ colliders and
compares them with the LHC. Noted specifically are examples where the energy
precision (E) or flavour non-universality (F) would be advantageous for a
$\mu^+\mu^-$ collider, and where the availability of $e\gamma$ and/or
$\gamma\gamma$ collisions ($\gamma$) or beam polarization (P) would favour an
$e^+e^-$ collider. It should also be commented that the experimental
environment
at a high-energy $\mu^+\mu^-$ collider is likely to be far more difficult
than a
CLIC. There is no way to prevent off-momentum $\mu^\pm$ passing through the
detector, though it should be possible to shield out the $e^\pm$ from $\mu^\pm$
decays.
\begin{table}
\caption{\it A comparison of some of the capabilities of high-energy
colliders, including the LHC, a second-generation linear $e^+ e^-$
collider and a $\mu^+ \mu^-$ collider at the high-energy frontier. for the
latter two cases, we note instances where photon beams ($\gamma$),
polarization (P), flavour non-universality (F) and energy calibration and
resolution (E) might be advantages.} 
\begin{center} \begin{tabular}{|l|l|l|l|} \hline
Physics topics & LHC & $e^+e^-$ & $\mu^+\mu^-$ \\ \hline
Supersymmetry  &&& \\
~~Heavy Higgses H, A & X? & ?:$\gamma$ & Y:F,E \\
~~Sfermions & $\tilde q$ & $\tilde\ell$& $\tilde\ell$: F \\
~~Charginos & X? & Y: P & Y: F,E \\
~~$R$ Violation & $\tilde q$ decays & $\lambda_{1ij}$ & $\lambda_{2ij}$: F,E \\
~~SUSY breaking & some & more & detail: F,E \\ \hline
Strong Higgs sector &&& \\
~Continuum & $\lappeq$ 1.5 TeV & $\lappeq$ 2 TeV & $\lappeq$ 2 TeV \\
~~Resonances & scalar, vector & vector, scalar & vector (E), scalar (F) \\ \hline
Extra dimensions &&& \\
~~Missing energy & large $E_T$ & Y & Y : E? \\
~~Resonances& $q^*, g^*$ & $\gamma^*, Z^*, e^*$ & $\gamma^*, Z^*, \mu^*$: E \\ \hline
\end{tabular}
\end{center}
\end{table}

The biggest obstacle to obtaining high energies in $\mu^+\mu^-$ colliders
may be
$\nu$ radiation~\cite{nurad}, which may even become a health hazard at
$E_{cm} \gappeq$
3 TeV.
Neutrinos will radiate in all directions in the plane of the collider ring,
with
particular concentrations in the directions of any straight sections. In
contrast
to a $\nu$ factory, where these should be as long as possible relative to the
arcs,  in a high-energy $\mu^+\mu^-$ collider one would like them to be as
short
as possible. Other strategies for reducing the $\nu$ radiation hazard include
burying it in a deeper tunnel, learning to be more efficient in using muons to
produce collider luminosity, and subtle choices of `ring' geometry.

\section{Present Accelerator R\&D Activities at CERN}

In its current Medium-Term Plan,
the present CERN management has expanded accelerator R\&D activities
at CERN, including work on both linear colliders and high-intensity
proton sources. A larger fraction of the resources available will be
directed
towards CLIC. As discussed here by Delahaye~\cite{CLIC}, it is hoped to
continue the
previous
successful studies with two successive stages, CTF3 and CLIC1, before
reaching a
stage (after 2008) when CLIC could be built. In parallel, a working group has
recently been charged to map out a strategy for R\&D towards a $\nu$ factory,
including studies of the proton driver, targetry, $\pi$ capture, $\mu$
cooling and
acceleration~\cite{NFWG}.

As a first step, four specific activities have been proposed~\cite{MUG}:

$\bullet$
an experiment to measure $\pi$ production,

$\bullet$
tests of RF cavities in a radiation environment
with a strong magnetic field,

$\bullet$
measurements of wide-angle muon  scattering,
with a view to better modelling of cooling channels, and

$\bullet$
target studies.

In parallel to these accelerator R\&D activities, there are physics study
groups
for $\nu$ beams and detectors (concentrating on oscillation
experiments)~\cite{Dydak}, on
$\mu^+\mu^-$ colliders~\cite{Janot}, and
on other possible physics with stopped muons, $\nu$ scattering,
etc.~\cite{EG}.
Simultaneously, there are parallel accelerator and physics working groups at
FNAL~\cite{FNAL},
the NSF has commissioned an Expression of Interest for  R\&D towards a
$\nu$
factory~\cite{NSF}, and a second  international workshop is scheduled for
Monterey in May
2000~\cite{Monterey}.

\section{Prospects}

As reviewed in this talk, there are clearly several interesting options for
possible accelerators at CERN beyond the LHC, some of which are being studied
quite actively, with CLIC as a default option~\cite{CLIC}. The relative
priorities of the various options before CERN will depend on project
developments elsewhere as well as on physics developments. In the coming years,
there will clearly need to be mutual understanding and coordination between
accelerator laboratories in different regions of the world, so as to arrive
at a
suitable distribution of projects. There is already worldwide interest in
linear
$e^+e^-$ colliders, and active discussion of different projects. In a few
years'
time, a similar stage may be reached for $\nu$ factories. Global
coordination on R\&D is already underway, and a similarly
cooperative approach to siting optimization would be desirable.
Hopefully, we will
eventually see a `World-Wide Neutrino Web' consisting of an intense
proton
source in one region feeding neutrino beams to detectors in different
regions --
a true World Laboratory for $\nu$ Physics. A Eurocentric
vision of this concept is shown in Fig. 16: see~\cite{NSF} for two
competing
American visions. 

CERN is preparing actively~\cite{CLIC,BS,NFWG,MUG} to play
whatever role seems most interesting and
appropriate in the generation of accelerators following
the LHC.

\begin{figure}
\begin{center}
\mbox{\epsfig{file=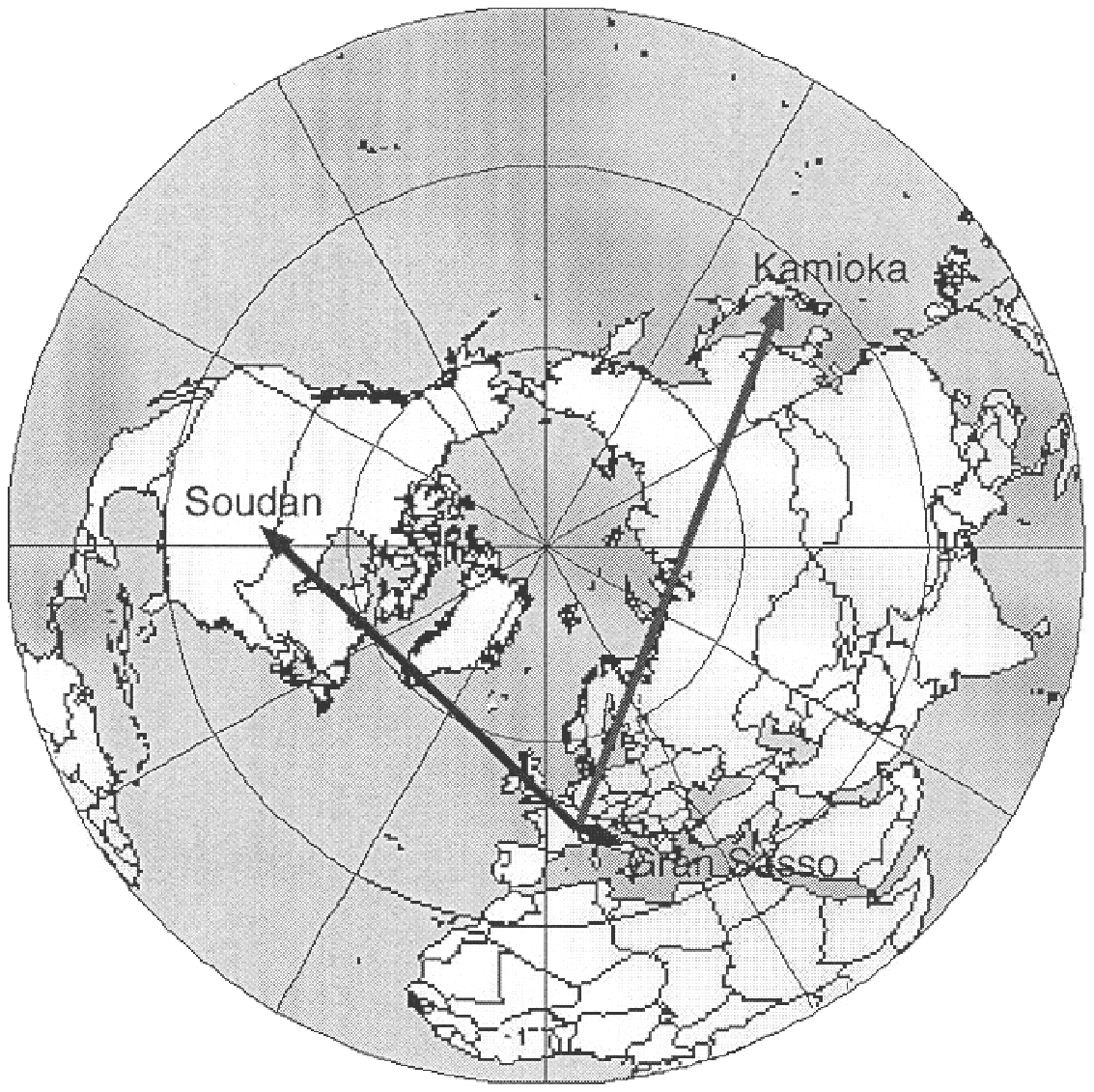,width=8cm}}
\end{center}
\captive{\it A Eurocentric view of the possible World-Wide Neutrino
Web, showing a source at CERN sending $\nu$ beams to the Gran Sasso
laboratory, the Soudan mine and Super-Kamiokande.}
\end{figure}

\newpage
{\bf Acknowledgements}
{~~}\\
\noindent
It is a pleasure to thank many colleagues for discussions on the
issues raised here, in particular Bruno Autin, Alain Blondel,
Friedrich Dydak, Belen Gavela, Helmut Haseroth, Eberhard Keil,
Luciano Maiani and
Gigi Rolandi. However, I take personal responsibility for the views
expressed here.

\end{document}